\documentclass[10pt,journal,epsfig]{IEEEtran}

\usepackage[dvips]{graphicx}
\usepackage{graphicx}
\usepackage{amssymb}
\usepackage{cite}
\usepackage{subfigure}
\usepackage{amsmath}
\usepackage{algorithm}
\usepackage{algorithmic}
\usepackage{multirow}
\usepackage{xcolor}
\usepackage{etoolbox}
\usepackage{upgreek}
\usepackage{tikz}
\usepackage{lipsum}

\makeatletter \patchcmd{\@makecaption}{\scshape}{}{}{}\makeatother

\IEEEoverridecommandlockouts

\newcommand\copyrighttext{%
  \footnotesize \textcopyright 2021 IEEE. Personal use of this material is permitted.
  Permission from IEEE must be obtained for all other uses, in any current or future
  media, including reprinting/republishing this material for advertising or promotional
  purposes, creating new collective works, for resale or redistribution to servers or
  lists, or reuse of any copyrighted component of this work in other works.
  }
\newcommand\copyrightnotice{%
\begin{tikzpicture}[remember picture,overlay]
\node[anchor=south,yshift=10pt] at (current page.south) {\fbox{\parbox{\dimexpr\textwidth-\fboxsep-\fboxrule\relax}{\copyrighttext}}};
\end{tikzpicture}%
}

\begin{document}
%% copyright stat-1

%% copyright-end-1

\title{Recent Advances on Sub-Nyquist Sampling-Based Wideband Spectrum Sensing}

\author{Jun Fang, Bin Wang, Hongbin Li,
~\IEEEmembership{Fellow,~IEEE}, and Ying-Chang Liang,
~\IEEEmembership{Fellow,~IEEE}
\thanks{Jun Fang, Bin Wang and Ying-Chang Liang are with the National Key Laboratory
of Science and Technology on Communications, University of
Electronic Science and Technology of China, Chengdu 611731, China,
Email: JunFang@uestc.edu.cn; 201611260116@std.uestc.edu.cn;
liangyc@ieee.org}
\thanks{Hongbin Li is with the Department of Electrical and Computer Engineering,
Stevens Institute of Technology, Hoboken, NJ 07030, USA, E-mail:
Hongbin.Li@stevens.edu}
%\thanks{This is an open call article which is recommended for
%acceptance by Dr. Xiqing Liu.}
%\thanks{This work was supported in part by the National Science
%Foundation of China under Grants 61934008 and 61871091.}
}

\maketitle
\copyrightnotice

\begin{abstract}
Cognitive radio (CR) is a promising technology enabling efficient
utilization of the spectrum resource for future wireless systems.
As future CR networks are envisioned to operate over a wide
frequency range, advanced wideband spectrum sensing (WBSS) capable
of quickly and reliably detecting idle spectrum bands across a
wide frequency span is essential. In this article, we provide an
overview of recent advances on sub-Nyquist sampling-based WBSS
techniques, including compressed sensing-based methods and
compressive covariance sensing-based methods. An elaborate
discussion of the pros and cons of each approach is presented,
along with some challenging issues for future research. A
comparative study suggests that the compressive covariance
sensing-based approach offers a more competitive solution for
reliable real-time WBSS.
\end{abstract}

\section{Introduction}
The unprecedented progress of wireless communications has spurred
an extensive deployment of wireless devices, through which a
tremendous amount of wireless data needs to be transmitted at all
times. It is predicted that mobile data traffic will grow by about
$1000$ times in the next decade. Such a proliferated increase
magnifies the scarcity of radio frequency (RF) spectrum resources.
In the future networks, it is prospected that the wireless system
should achieve a significant increase in capacity, spectrum
efficiency, and energy efficiency by at least 10 times up to 1000
times \cite{AndrewsBuzzi14}. To realize these grand visions, on
one hand, new spectrum resources should be explored, while on the
other hand, current spectrum entails to be more efficiently
utilized. As such, a spectrum coordination system that is capable
of easing the mobile traffic tension is imperatively required.

In traditional spectrum-management approaches, the frequency bands
are exclusively allocated to primary users (PU) or licensed users.
However, it has been reported by the Federal Communications
Commission (FCC) that the spectrum localized in time and space is
highly underutilized. Such an underutilization has propelled a
considerable research interest in cognitive radio (CR)
\cite{MitolaMaguire99}, which is a tempting paradigm that
automatically senses a PU's surrounding RF environment to provide
opportunistic access to the idle spectrum. A key component in CR
is spectrum sensing, which aims to detect the presence of
incumbent signals and achieve interference avoidance, because PUs
have no obligation to share or abalienate their spectrum. So far
most studies are confined to perform sensing over a narrow band,
which is also referred to as narrowband spectrum sensing (NBSS).
Nevertheless, from a long-term perspective, it is of vital
importance to perform sensing over a wide frequency range in order
to provide more flexible dynamic access and achieve a higher
opportunistic throughput. The need is evidenced in the 2012
report, ``Realizing the Full Potential of Government-held Spectrum
to Spur Economic Growth'', authored by the US President's Council
of Advisors on Science and Technology (PCAST), which advocated the
concept of ``shared-use spectrum superhighways'' and recommended
to share a $1$ gigahertz (GHz) wide federal government spectrum
ranging from $2.7$ to $3.7$ GHz with non-government entities for
open access. Clearly, advanced WBSS techniques that can quickly
detect idle spectrum over a wide frequency span is an enabling
technology to realize the concept of shared-use spectrum
superhighways. Analogously, applications like spectrum
aggregation, which integrates intermittent spectrum slots to
enhance user experiences, also need to oversee a wide spectrum
range.

According to the classic Nyquist--Shannon theorem, one has to
sample the signal at the Nyquist rate in order to realize
real-time WBSS. When the spectrum to be sensed is very wide, this
may bring a challenge to hardware implementation as high-speed
analog-digital converters (ADCs) are energy-intensive and too
costly for practical systems. Thus, sub-Nyquist sampling-based
techniques which have the potential to accomplish real-time WBSS
in a cost-effective and energy-efficient way are highly desirable
and have received much attention recently \cite{SunNallanathan13}.

In this article, we start with an introduction of the conventional
NBSS. Then we provide an overview of recent advances on
sub-Nyquist sampling-based WBSS methods, which can be classified
into two broad categories: compressed sensing-based approach and
compressive covariance sensing-based approach. We discuss the pros
and cons of these two approaches, and review related
state-of-the-art techniques. Future research challenges are
subsequently highlighted, followed by concluding remarks.

\section{Narrowband Spectrum Sensing}
The term narrowband means that the range of the frequency band is
sufficiently small such that the whole frequency band is either
occupied by the PU or available for opportunistic access. NBSS
usually boils down to a binary hypothesis testing problem. When
the PU's signal waveform is known \emph{a priori}, the matched
filter detection (MFD), which correlates the received signal with
the known waveform, is statistically optimal for this problem. The
MFD, however, not only requires the knowledge of the PU's waveform
but also an accurate synchronization between the PU and the
secondary user (SU), which limits its practical applications. When
little is known about the PU's signal, energy detection (ED) which
distinguishes the PU's presence and absence by measuring the
energy of the received signal can be adopted for spectrum sensing.
The energy detector is easy to implement and also computationally
cheap. Nevertheless, it is sensitive to the noise uncertainty and
does not provide satisfactory performance in the low
signal-to-noise (SNR) regime.

Over the past decade, accompanying the development of the
multiple-in-multiple-out (MIMO) technology, multi-antenna spectrum
sensing has become the center of interest in the research of
narrowband spectrum sensing. By exploiting the spatial correlation
among different receive antennas, a variety of eigenvalue-based
spectrum sensing algorithms were proposed. For example, the ratio
of the maximum to the minimum eigenvalue of the receive sample
covariance matrix is employed in \cite{ZengLiang09} as the test
statistic for spectrum sensing. More recently, the great success
of deep learning in a variety of learning tasks has inspired
researchers to use it as an effective tool to devise model-free
spectrum sensing algorithms \cite{LiuWang19}. In contrast to
traditional spectrum sensing algorithms that are model-based and
perform detection based on the distribution of the received signal
samples, deep learning-based methods are data-driven with the test
statistics or features generated directly from signal samples.
When sufficient training is available, deep learning-based
algorithms can provide superior spectrum detection performance
thanks to the powerful ability of neural networks in learning
features out of signal samples \cite{LiuWang19}.

\begin{table*}[t]
    \centering
    \subfigure{\includegraphics[width=7in]{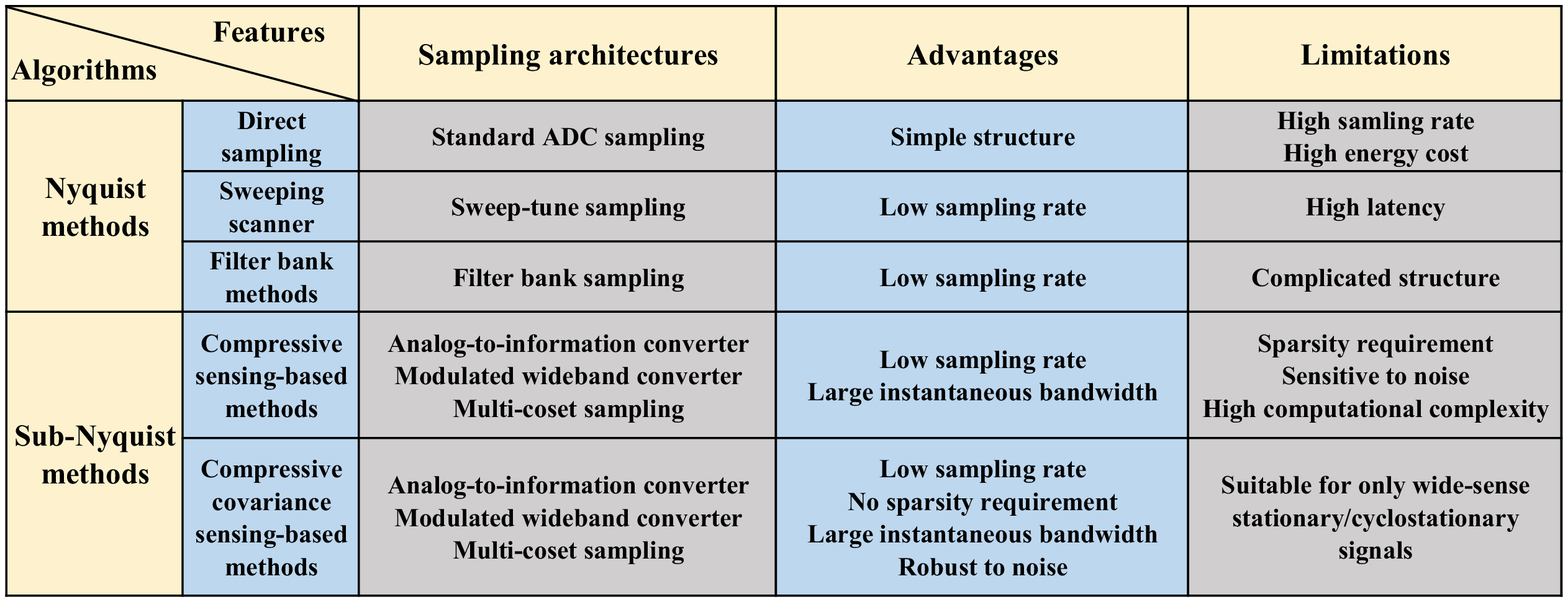}} \hfil
    \caption{Summary of sampling architectures, advantages and
    limitations
    of existing WBSS techniques.}
    \label{table1}
\end{table*}

\section{Wideband Spectrum Sensing}
A straightforward approach for WBSS is to use a sufficiently
high-speed ADC to meet the Nyquist--Shannon theorem. This
approach, however, becomes prohibitively expensive or even
infeasible when the frequency range is excessively wide, say, of a
few GHz. With current hardware technologies, high-speed,
high-precision ADCs are either not available, or too costly and
power hungry. For example, some high-end ADCs like AD9213 can
provide ultrafast sampling rates beyond 10 GSPS, but have a high
power consumption up to a few Watts and cost about several
thousands of US dollars. In addition, a high sampling rate
generates a large amount of data and imposes a significant
challenge on subsequent data storage, processing and transmission.
To address these difficulties, an alternative method is to
sequentially scan the frequency band of interest by resorting to
the frequency mixing (superheterodyne) technique. The
superheterodyne architecture, however, relies on high-quality RF
components that need to be widely-tunable and thus can hardly be
implemented on chip. Moreover, when the spectrum under monitoring
reaches gigahertz, such a scanning scheme may incur a sensing
latency from tens of milliseconds to the order of seconds, thus
preventing the receiver from exploiting spectral opportunities in
a more efficient way. Another solution is to uniformly divide the
wideband signal into multi-narrowband signals by a number of
parallel frequency-shifted bandpass filters, and each narrowband
signal can be sensed at the Nyquist rate. Although the sensing
latency is mitigated, the success of this approach comes at the
expense of the complex structure of parallel filter banks as well
as requiring a large number of RF components.

Recently, as a new paradigm for data sampling and acquisition,
compressed sensing has inspired a wide interest in sub-Nyquist
sampling-based WBSS. Many efforts have been made toward this
direction and a plethora of sub-Nyquist-based spectrum sensing
methods have been developed. These methods can be generally
classified into two categories: compressed sensing-based approach
and compressive covariance sensing-based approach, depending on
what kind of information is extracted from sub-Nyquist data
samples. In the sequel, we provide a review of each approach and
discuss their respective pros and cons. Also, for clarity, a
concise overview of some key aspects of existing WBSS techniques
is summarized in Table \ref{table1}.

\subsection{Compressed Sensing-Based Methods}
Compressed sensing-based methods are motivated by the fact that
the spectrum localized in time and space is in general severely
underutilized. Utilizing the sparsity structure of the frequency
domain, WBSS can be formulated as a compressed sensing problem,
which aims to reconstruct the wideband analog signal from
sub-Nyquist data samples. To bridge the analog and the digital
domain, compressed sampling schemes including
analog-to-information converter (AIC) \cite{ArianandaLeus12},
modulated wideband converter (MWC) \cite{MishaliEldar11a}, and
multi-coset sampling (MCS), were proposed to convert the wideband
analog signal into digital compressive measurements. Although
different in hardware implementation, their common objective is to
obtain linearly compressive measurements of the Nyquist data
samples of the analog wideband signal.

\begin{figure*}[t]
    \centering
    \includegraphics[width=16cm]{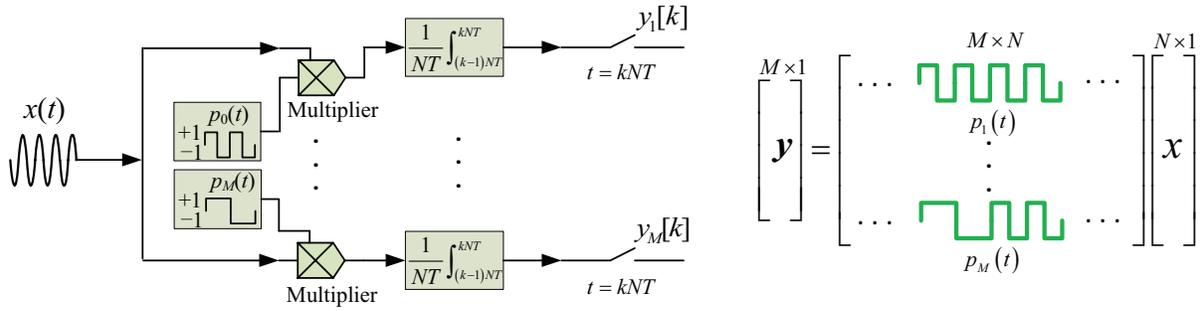}
    \caption{Compressed sampling architecture: AIC, where $M$ denotes the number of
sampling branches, $N$ represents the downsampling factor, $T$ is
the Nyquist sampling interval, and $p_m(t)$ denotes the PN
sequence.}
    \label{fig1}
\end{figure*}

Consider the AIC as an example to illustrate the basic idea of the
compressed sensing-based WBSS methods. The AIC sampling scheme
encompasses a set of parallel sampling branches. In each branch,
the wideband analog signal is modulated by a periodic pseudo-noise
(PN) sequence which takes constant values within each Nyquist
interval. Specifically, the period of the PN sequence is an
integer multiple of the Nyquist interval, and the ratio of the PN
sequence's period to the Nyquist interval is called as the
downsampling factor, which is a positive integer of user's choice.
After the PN sequence modulation, the signal is then passed
through an integrate-and-dump device whose period is the same as
that of the PN sequence. At last, the analog signal is sampled by
an ADC whose sampling rate is equal to the Nyquist rate divided by
the downsampling factor. Note that multiple branches are sampled
synchronously. Let $\boldsymbol{y}$ be a sub-Nyquist sample vector
obtained by stacking the sampled signals of all branches. The
signal acquisition process can thus be described as an
underdetermined system of linear equations (see Fig. \ref{fig1}):
\begin{align}
\boldsymbol{y}= \boldsymbol{C}\boldsymbol{x} \label{eqn1}
\end{align}
where $\boldsymbol{C}$ is a matrix constructed by PN sequences,
and $\boldsymbol{x}$ consists of Nyquist samples collected within
a period of the PN sequence. Since the spectrum under monitoring
is underutilized, it has a sparse representation in the DFT
domain. Thus the Nyquist data sample vector can be reconstructed
from sub-Nyquist samples via compressed sensing techniques,
provided that the restricted isometric property is satisfied
\cite{TroppWright10}.

Some encouraging results have been reported on hardware
implementation of compressed sensing-based wideband signal
receivers. The first compressed sampling hardware was reported in
\cite{MishaliEldar11a}, where a wideband receiver prototype based
on the MWC sampling scheme was developed, which can deal with 2GHz
Nyquist-rate input signals with a total sampling rate of 280
megahertz (MHz). Later in \cite{YazicigilHaque15}, a WBSS detector
was developed based on a quadrature AIC (QAIC). The detector,
implemented in 65 nm CMOS, is capable of quickly detecting the
presence of strong interference signals over the PCAST Band.

Although these compressed sensing-based approaches are of
theoretical and practical values, they suffer several severe
drawbacks. Firstly, sparse signal recovery via optimization
methods or other heuristic methods incurs a high computational
complexity that grows quadratically with the problem size. Table
\ref{table2} provides a summary of the computational complexities
of both compressed sensing-based and compressive covariance
sensing-based WBSS methods, which shows that compressed
sensing-based methods generally have a higher complexity than
compressive covariance sensing-based methods. Secondly, a
practical system is inevitably subject to measurement noise.
Sparse signal recovery algorithms, however, usually require a
moderately high SNR to achieve satisfactory performance. It was
reported in \cite{MishaliEldar11a} that to ensure correct recovery
of the spectrum support, an SNR above 15dB is needed. Such a high
SNR requirement results in a low receiver sensitivity, which could
potentially lead to a miss of some strong signals that should have
been detected for interference avoidance/management.

In \cite{SunNallanathan16}, a multirate sub-Nyquist sampling
scheme was proposed to realize WBSS. The proposed approach does
not intend to recover the wideband analog signal. The rationale
behind the work is to employ multiple sub-Nyquist sampling rates
to sample a wideband analog signal. When the spectral occupancy
rate is sufficiently low and these sub-Nyquist sampling rates
satisfy a certain condition, the probability that different
sub-Nyquist samplers have aliased frequencies on the same
frequency bin is small. Thus the spectral occupancy can be
estimated by fusing detection results from different sub-Nyquist
samplers. Although not aiming to directly reconstruct the wideband
signal, this method can still be considered as a compressed
sensing-based approach with the objective of recovering the
support set of a sparse signal.

\begin{table*}[t]
    \centering
    \subfigure{\includegraphics[width=16cm]{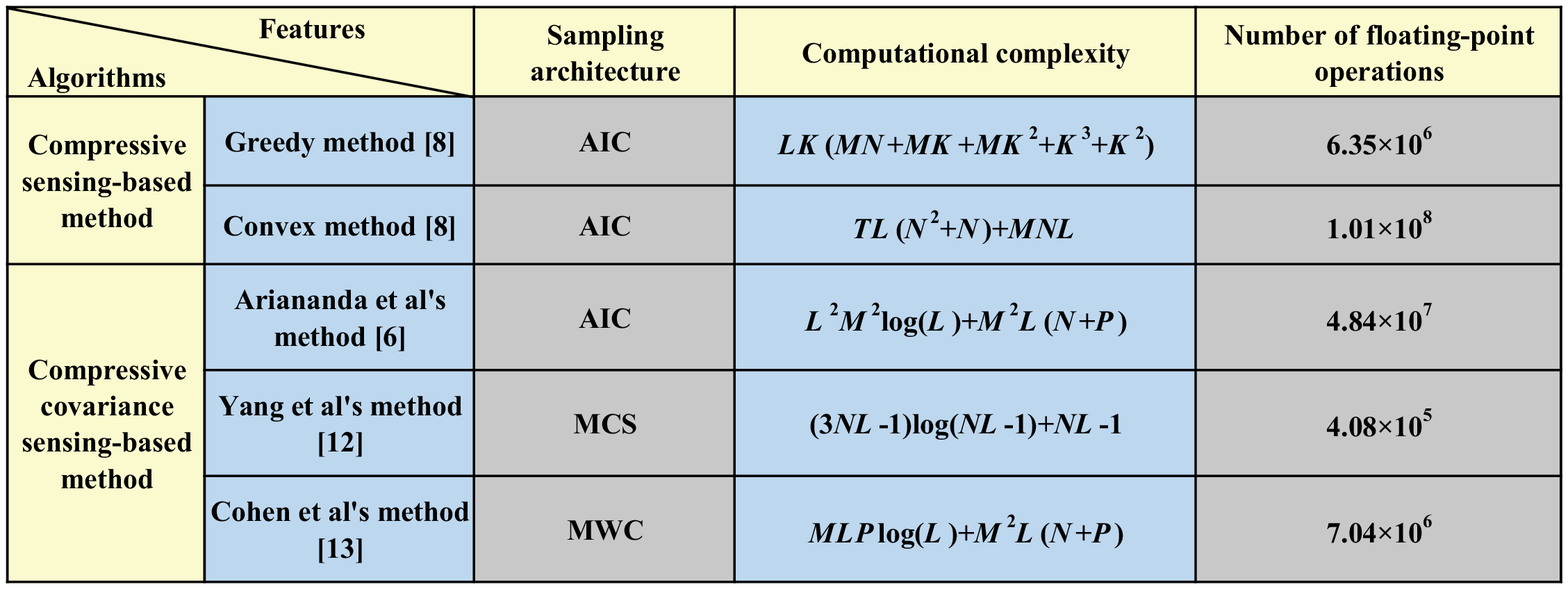}} \hfil
    \caption{Summary of computational complexity of existing WBSS
    techniques, where $M$ denotes the number of
sampling branches, $N$ represents the downsampling factor, $L$ is
the number of samples (in time) chosen to attain a specified
spectrum resolution, $K$ denotes the sparsity level in an
$N$-dimensional sparse signal, $T$ stands for the number of
iterations required for the convex method, and $P$ represents the
number of samples used to calculate the correlation matrix of
sub-Nyquist signals. The number of floating-point operations is
calculated based on the following setup: Suppose we aim to sense a
frequency band of 1GHz with a spectrum resolution of 100kHz. The
spectrum occupancy ratio is assumed to be 10\%. The sensing system
has $25$ sampling channels and the downsampling factor is set to
$100$. We choose $L$ to be $100$ to achieve the desired spectrum
resolution, and $P$ and $T$ are respectively set as $10$ and
$100$.}
    \label{table2}
\end{table*}

\subsection{Compressive Covariance Sensing-Based Methods}
To overcome the difficulties of compressed sensing-based methods,
some recent works, e.g. \cite{RomeroAriananda15}, proposed to
reconstruct the covariance matrix (equivalently, power spectrum)
of the wideband signal, instead of the signal itself, from
sub-Nyquist data samples. This class of methods are referred to as
the compressive covariance sensing-based or the compressed power
spectrum estimation approach. Compressive covariance sensing-based
approaches use the same compressed sampling architectures as those
of compressed sensing-based methods. Note that WBSS can be
accomplished simply based on the power spectrum of a wideband
signal. Also, recovering the power spectrum can bring in some
noteworthy advantages.

Firstly, compressive covariance sensing-based methods can
reconstruct the power spectrum reliably even in a low SNR
environment, since the white noise is averaged out or effectively
suppressed in the second-order statistics. Empirical results
reported later in this section show that compressed power spectrum
estimation methods can achieve reliable spectrum sensing
performance even when the SNR is below 0dB. Secondly, compressive
covariance sensing-based methods are in general computationally
more efficient than compressed sensing-based methods as they do
not require solving a computationally costly sparse recovery
problem. For example, the fast Fourier transform (FFT)-based
method proposed in \cite{YangFang20} requires several FFTs to
reconstruct the power spectrum from sub-Nyquist data samples,
making it possible to perform real-time WBSS using highly mature
commercial solutions such as field-programmable gate array (FPGA).
Lastly, unlike compressed sensing-based methods, compressive
covariance sensing-based methods do not need to impose any
sparsity requirement on the frequency domain. This is because the
covariance matrix to be recovered has a Toeplitz structure, which
can be utilized to significantly reduce the number of unknowns and
ensure an overdetermined linear system without requiring the
sparsity of the wideband signal. Such a merit enables the CR to
efficiently utilize spectrum fragments in a crowded frequency
band.

We now provide an overview of state-of-the-art compressive
covariance sensing-based WBSS methods. There are two different
approaches to compressive covariance sensing-based WBSS, namely, a
time-domain approach and a frequency-domain approach. The
time-domain approach tackles the power spectrum estimation problem
from a time-domain perspective, and aims to reconstruct the
autocorrelation of a wide-sense stationary signal. Again, let us
take the AIC as an example. Based on the linear relationship
(\ref{eqn1}), it is clear that the covariance matrix of the
sub-Nyquist sample vector can be expressed in terms of the
covariance matrix of the Nyquist sample vector as:
\begin{align}
\boldsymbol{R}_y=\boldsymbol{C}\boldsymbol{R}_x \boldsymbol{C}^T
\end{align}
Due to the wide-sense stationarity, the covariance matrix of the
Nyquist sample vector is a Toeplitz matrix, whose intrinsic degree
of freedom is much smaller than its natural dimension. For this
reason, reconstruction of the autocorrelation of the wide-sense
stationary signal from the autocorrelation of the sub-Nyquist
samples reduces to solving an over-determined linear system of
equations, as long as the PN sequences are properly designed and a
sufficient number of sampling channels are deployed
\cite{ArianandaLeus12}.

\begin{figure*}[t]
    \centering
    \includegraphics[width=16cm]{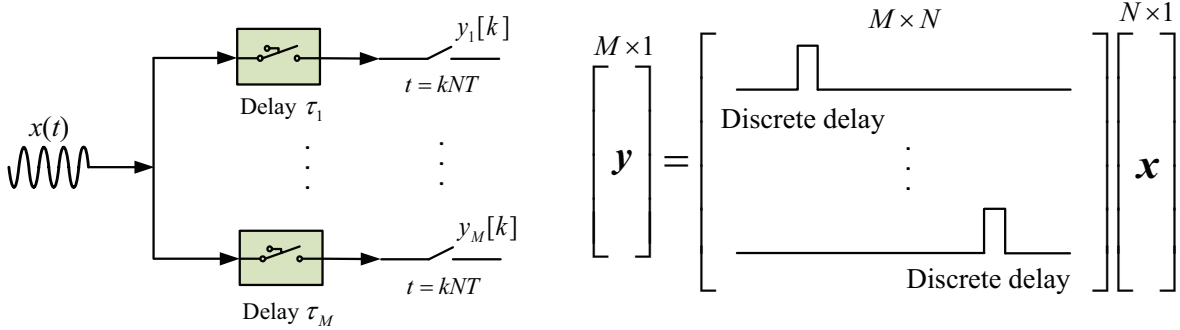}
    \caption{Compressed sampling architecture: MCS.}
    \label{fig2}
\end{figure*}

For the MCS scheme, the linear relationship between the
sub-Nyquist sample vector's covariance matrix and the Nyquist
sample vector's covariance matrix can be similarly established,
except that $\boldsymbol{C}$ is a selection matrix obtained via
extracting a portion of rows from an identity matrix (see Fig.
\ref{fig2}). The indices of those nonzero elements in rows of
$\boldsymbol{C}$ correspond to the time delays incurred by
different sampling branches. The condition for recovering the
autocorrelation of the wide-sense stationary signal from the
sub-Nyquist sample vector's covariance matrix was thoroughly
investigated in \cite{ArianandaLeus12,YangFang20}, where it is
shown that the recoverability is guaranteed if the time delay set
is a circular sparse ruler of a particular length. A circular
sparse ruler of length-$Q$ is a set of integer marks $\{a_m\}$
($0\leq a_m\leq Q$) such that it can measure all integers from 0
to $Q$ in a modular fashion. The problem of finding a smallest
number of marks (i.e. branches) such that we can construct a
circular sparse ruler of a specified length, also referred to as
minimal circular sparse ruler, is a combinatorial problem that has
attracted much attention in the field of sensor array processing
\cite{RomeroAriananda15}.

\begin{figure*}[t]
    \centering
    \includegraphics[width=16cm]{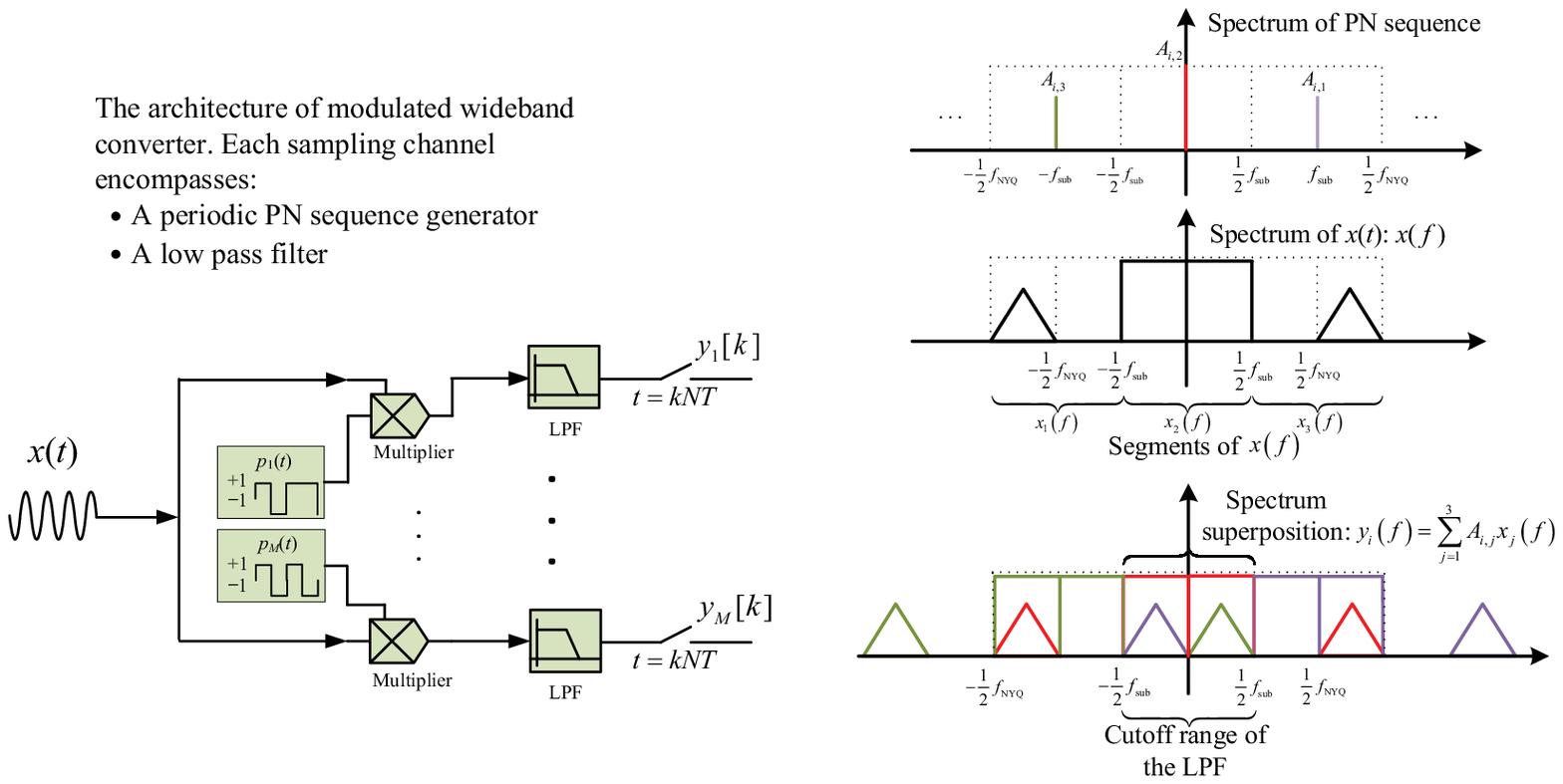}
    \caption{Compressed sampling architecture: MWC and an
    illustrative example of the spectrum of the low-pass filtered signal.}
    \label{fig4}
\end{figure*}

The compressed power spectrum estimation problem was also
addressed from the frequency-domain perspective. In
\cite{CohenEldar14}, authors proposed a frequency-domain approach
based on the MWC sampling scheme. The MWC architecture resembles
the AIC architecture except that the integrate-and-dump device is
replaced by a low-pass filter. Since the analog signal is
multiplied by a periodic PN sequence, the spectrum of the PN
modulated signal at each branch is a convolution of the spectrum
of the original signal and a train of impulses, in other words, it
is a weighted superposition of the original spectrum and its
frequency-shifted versions. After going through the low-pass
filter, the resulting spectrum is a superposition of a number of
equal-width segments of the spectrum of the original signal. To
facilitate readers' understanding, an illustrative example of this
linear relationship is provided in Fig. \ref{fig4}. In summary,
this sub-Nyquist sampling process can be expressed as an
underdetermined system of linear equations which characterize the
relationship between the Fourier transform of the low-pass
filtered signal and that of the original analog signal. As the
Fourier transform of a wide-sense stationary process is white
noise, the autocorrelation of the Fourier transform of the
original signal exhibits a diagonal structure. Thus reconstructing
the power spectrum of the original signal from the autocorrelation
of the Fourier transform of the low-pass filtered signal can be
cast into an over-determined inverse problem and readily solved.

Many man-made signals such as digital communication signals are
cyclostationary instead of wide-sense stationary because their
covariance function is periodic. In fact, wide-sense stationarity
is a special case of cyclostationarity since the former can be
regarded as a cyclostationary signal with a cyclic period of one.
It was shown in \cite{YangFang20} that when it comes to the
cyclostationary signal, compressive covariance sensing methods
based on the wide-sense stationarity assumption render an estimate
of the average power spectrum, i.e. the cyclic spectrum of the
cyclostationary signal at the zeroth cyclic frequency. To
explicitly accommodate cyclostationary signals, Tian and Tafesse
\cite{TianTafesse12} presented the first attempt to reconstruct
the cyclic spectrum from sub-Nyquist data samples. As opposed to
the stationary case, the Nyquist sample vector's covariance matrix
is no longer a Toeplitz matrix. Thus the linear relationship
between the Nyquist sample vector's covariance matrix and the
sub-Nyquist sample vetor's covariance matrix is usually
under-determined. However, it is observed that the cyclic spectrum
of a cyclostationary signal is sparse. Such a prior knowledge can
be incorporated to convert cyclic spectrum recovery into a sparse
recovery problem \cite{TianTafesse12}. From another point of view,
although the Nyquist sample vector's covariance matrix is
non-Toeplitz, it is block-Toeplitz because of the periodicity of
the covariance function. Inspired by this, \cite{ArianandaLeus14}
proposed to divide the vector of Nyquist samples into a number of
block vectors, where the length of each block is an integer
multiple of the cyclic period. Each block is then sampled by an
individual multi-coset sampling module to collect its associated
sub-Nyquist samples. By utilizing the block-Toeplitz structure,
the linear relationship between the autocorrelation of the
wideband signal and that of sub-Nyquist samples becomes
over-determined, which helps avoid the computationally costly
sparse recovery process.

\begin{figure}[!t]
\centering
\includegraphics[width=8.5cm]{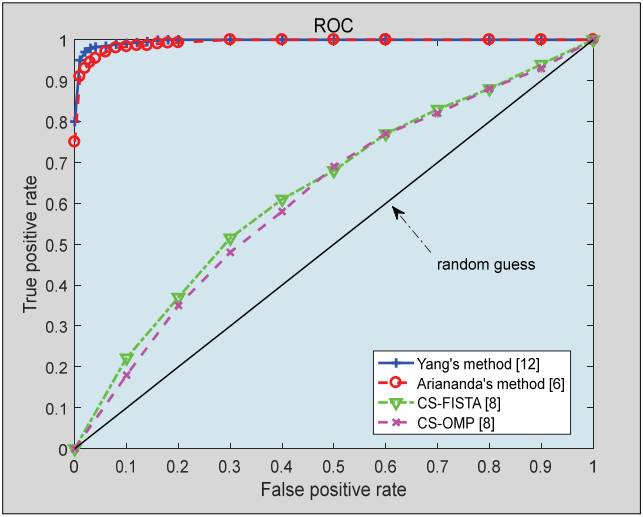}
\caption{Detection performance attained by respective methods. }
\label{fig11}
\end{figure}

In Fig. \ref{fig11}, we plot the ROC curves for both the
compressed sensing-based methods and compressive covariance
sensing-based methods, where the SNR is set to -10dB. We consider
sensing a frequency range of [0,1]GHz using a multicoset sampling
scheme with 8 sampling channels. The sampling rate for each
channel is 80MHz. We see that compressive covariance sensing-based
methods can render reliable spectrum sensing performance in the
low SNR regime, whereas the compressed sensing-based methods which
aim to recover the wideband analog signal suffer a considerable
amount of performance degradation.

\section{Open Research Challenges}
The research on sub-Nyquist sampling-based WBSS has witnessed
enormous progress over the past few years. Nevertheless, extensive
work is needed to obtain a more comprehensive understanding of
related theoretical aspects and to implement this technique in
practical systems. Some possible future directions are discussed
in this section.

\subsection{Innovative Sub-Nyquist Sampling Architectures}
Existing sub-Nyquist sampling architectures have some drawbacks. A
major difficulty of the AIC and MWC sampling schemes lies in that
the flipping rate of each PN sequence should be at least the same
as the Nyqusit rate. However, the high rate PN sequence generator
is expensive and also is the most power hungry block of AIC and
MWC. Moreover, a timing synchronization module is required to
synchronize different PN sequences. Although the MCS architecture
is simpler than the AIC and MWC, a precise timing control is
required to ensure that the time delays are exactly integer
multiples of the Nyquist interval. Unfortunately, maintaining
precise time delays is rather difficult due to time shift
inaccuracies, particularly when the time delays are at the order
of the Nyquist interval. Also, since MCS needs to sample the
analog wideband signal directly, the ADCs used to implement the
MCS require a high analog bandwidth to ensure that the analog
wideband signal is not severely distorted before being sampled. In
this regard, there is a need to quantify and provide an in-depth
understanding of the impact of the hardware imperfections on the
performance of practical systems. In addition, innovative
sub-Nyquist sampling architectures that can overcome those
drawbacks are highly desirable.

\subsection{Cooperative Wideband Spectrum Sensing}
Currently, the research on WBSS mainly focuses on the single CR
scenario. Due to shadowing or multipath fading, the received
signal at the CR may suffer severe degradation, leading to an
unreliable detection result. It is therefore necessary to study
cooperative schemes that encompass a networked system of CRs. In
such a cooperative system, a number of CRs that spread over a
large area are connected to a data center. If different nodes can
be well synchronized, then cooperation can be performed at a
bottom level, where raw sub-Nyquist data samples are pulled
together at the date center for joint WBSS. Specifically, since
different nodes could share some common spectral components, this
spatial correlation may be utilized to devise a cooperative
sub-Nyquist sampling scheme to reduce the overall sampling rate
required for WBSS. Nevertheless, such a bottom-level fusion scheme
may involve a large amount of communication overhead, especially
when there is a large number of CRs. To address this issue,
cooperation can be performed at a higher level, i.e. nodes report
their initial detection results or signal statistics to the data
center, where signal processing techniques can be developed to
fuse these high-level information by incorporating the data
correlation observed by different nodes as well as node's location
information.

\subsection{Multi-Antenna-Assisted Wideband Spectrum Sensing}
Recently, there is a growing interest in sub-Nyquist
sampling-based algorithms for joint WBSS and direction-of-arrival
(DoA) estimation in a phased-array framework. Due to the emergence
of massive MIMO beamforming techniques, future electromagnetic
environments could exhibit directionality in a local area. In
particular, a beamformed signal is significant in only a limited
number of directions. This fact potentially enables a same
frequency band to be reused in the spatial domain. In this
perspective, joint WBSS and DoA estimation would allow the
spectrum resource to be more efficiently utilized. Moreover,
compared with the single-antenna receiver, multiple antennas which
exploit spatial diversity can effectively mitigate multipath
fading, thereby improving the reliability of spectrum sensing. In
the multi-antenna context, new sub-Nyquist sampling architectures
as well as methods for WBSS are worthy of future investigation.

\section{Concluding Remarks}
This article presented a survey of recent developments on WBSS. In
particular, our survey is emphasized on sub-Nyquist sampling-based
WBSS methods, as these methods have the potential to realize
real-time WBSS in a cost-effective and energy-efficient way. Based
on the underlying principles, we classified sub-Nyquist
sampling-based methods into two categories, namely, a compressed
sensing-based approach and a compressive covariance sensing-based
approach. State-of-the-art techniques in each category were
reviewed, along with an extensive discussion on the pros and cons
of these two different kinds of approaches. Finally, several open
research issues for sub-Nyquist sampling-based WBSS were
presented.

\end{document}